# An Empirical Study of the Relation Between Community Structure and Transitivity

Keziban Orman, Vincent Labatut, Hocine Cherifi

**Abstract** One of the most prominent properties in real-world networks is the presence of a community structure, i.e. dense and loosely interconnected groups of nodes called communities. In an attempt to better understand this concept, we study the relationship between the strength of the community structure and the network transitivity (or clustering coefficient). Although intuitively appealing, this analysis was not performed before. We adopt an approach based on random models to empirically study how one property varies depending on the other. It turns out the transitivity increases with the community structure strength, and is also affected by the distribution of the community sizes. Furthermore, increasing the transitivity also results in a stronger community structure. More surprisingly, if a very weak community structure causes almost zero transitivity, the opposite is not true and a network with a close to zero transitivity can still have a clearly defined community structure. Further analytical work is necessary to characterize the exact nature of the identified relationship.

## 1 Introduction

In a complex network, a *community* is a cohesive subset of nodes with denser inner links, relatively to the rest of the network [1]. The presence of such groups is a common feature in networks modeling different types of real-world systems, including biological, social, information or technological ones [2]. When a network takes the form of a set of interconnected communities, it is said to possess a *community structure*.

The presence of a community structure is presumably related to other topological properties of the network. Uncovering what causes a community structure to appear, and what its effects are, would be valuable for a better understanding of the complex networks structure and dynamics. In particular, it would allow im-

Keziban Orman
Galatasaray Uni versity & University of Burgundy, e-mail: korman@gsu.edu.tr

Vincent Labatut
Galatasaray University, e-mail: vlabatut@gsu.edu.tr

Hocine Cherifi
University of Burgundy, e-mail: hocine.cherifi@u-bourgogne.fr

proving or explaining the existing community detection methods, and provide tools to interpret the communities identified in real-world networks. This angle was adopted in a few studies, with different objectives and/or in different contexts.

Pastor-Satorras *et al.* [3] showed how the presence of a hierarchical community structure and a power law degree distribution are sufficient conditions to cause a high transitivity (also called clustering coefficient). For this matter, they defined a generative model implementing these properties and studied the obtained networks. Moreover, they derived a new use for the transitivity measure, by utilizing its distribution to characterize the network hierarchical structure. Clauset *et al.* [4] proposed a different hierarchical approach: they defined a parameterized hierarchical model which they fit to various real-world data. The obtained hierarchical structures possess various properties present in real-world networks, including being scale-free (power law distributed degree) and having a high transitivity. This seems to indicate the hierarchical structure alone is enough to get both a scale-free and highly transitive network. Lie & Hu [5] proposed a model able to generate networks with community structures of various strengths, and showed the transitivity of the resulting networks depend on this strength. They used their model to study the effect of community structure on the network epidemic threshold. Interestingly, the generated networks are neither scale-free nor have a hierarchical structure, which seems to indicate these are sufficient, but not necessary conditions. Wang and Qin [6] had the same objective, but used a different model. It is a mixture of Watts-Strogatz's small-world model [7] and Newman's community structure model [1]. It is therefore not hierarchical either, nor is it scale-free.

The previous studies intended at studying the effects of the community structure on some topological properties of interest. In the works by Jin *et. al* [8] and Boguñá *et. al* [9], the community structure is, on the contrary, a byproduct. The authors focused on social networks and designed their models as multi-agent systems mimicking social interaction. The generated networks turned out to possess some properties observed in real-world social networks, including hierarchical community structure and high transitivity. Interestingly, the degree is not power law-distributed in social networks, which seems to confirm the scale-free property is not a prerequisite to get highly transitive and/or community structured networks.

In this article, our goal is to study how transitivity and community structure can mutually affect each other in realistic networks. Contrarily to the first cited studies [3, 4], we consider non-hierarchical networks, since this property does not seem to be a necessary condition to the presence of a community structure. The obvious difference with studies [5, 6] is our focus on transitivity, which intuitively seems to be a good candidate to explain the presence of a community structure (cf. section 3 ). Another important difference is our aim of evaluating not only the effect of the community structure on this property, like in [3-6], but also the effect of the property on the community structure. Finally, we are not interested in the specific process resulting in the network structure, like in [8, 9], but rather in the general relationships between community structure and transitivity.

To study this relationship, we adopt an empirical approach based on several generative models. First, we use an existing model to generate realistic networks possessing a community structure with a controlled strength, [10] and study its transitivity. Second, an existing model [11] and a new model of our own are used to generate networks with a high or controlled transitivity, and we study the strength of their community structure. The rest of the document is structured as follows. In the next two sections, we review the notion of community structure and justify our choice of the transitivity as a property of interest relatively to its study. Section 4 is dedicated to the description of our methods, and more particularly the models we are using. We then present the results of our simulation and discuss the nature of the uncovered relationships in sections 5.1 and 5.2 . Finally, we conclude by highlighting our contributions and the possible extensions of our work.

## 2    Community Structure

The concept of community can be formally defined in several ways: mutually exclusive vs. overlapping, hierarchical vs. flat, local vs. global, etc. [12]. The nature of the community structure directly depends on the considered definition of a community. Independently from this choice, stating the presence or absence of a community structure is itself an ambiguous task. For this matter, one can clearly distinguish two extreme cases: on the one hand, the complete absence of any community structure (e.g. a complete network, in which all nodes are connected to each others), and on the other hand a perfect community structure (a network made up of several disconnected components). Between these two extremes lies a continuum of networks exhibiting community structures of various strengths. It makes therefore more sense to measure this strength rather than the presence or absence of a community structure.

In this article we selected the *modularity* [1] for this matter. It is certainly the most widely spread measure to assess the strength of a community structure. It is based on the numbers of intra- and inter-community links, and consists in comparing the proportion of intra-community links present in the network of interest, to the expectation of the same quantity for a randomly generated network of similar size and degree distribution. It is worth noticing some limits have been identified since the creation of this measure [12]. The most important seems to be its resolution limit, causing it to fail identifying communities considered as small relatively to the network size and community interconnection pattern [13]. However, we considered it to be sufficient for this exploratory work.

Let us note $e_{ij}$ the proportion of links connecting nodes in community $i$ to nodes in community $j$. Then the proportion of intra-community links for the whole network is $\sum_i e_{ii}$. Let us note $e_{i+} = e_{+i}$ the proportion of links connecting at least one node from community $i$. For the same community, Newman defines the expected number of inter-community links as $e_{i+}^2$, in a network whose links are distributed randomly. The modularity is therefore: $Q = \sum_i (e_{ii} - e_{i+}^2)$.

4## 3 Transitivity

The *transitivity* (also called *clustering coefficient*) of a network is the relative proportion of triangles among all connected triads it contains [14]: $C = n_\Delta/n_\Lambda$ where $n_\Delta$ and $n_\Lambda$ are the numbers of triangles and connected triads, respectively. A triangle is a set of three completely connected nodes, whereas a triad can be either a triangle, or a set of three nodes connected by only two links (instead of three). The transitivity can be interpreted as the probability of finding a direct connection between two nodes having a common neighbor. The measure therefore ranges from 0 to 1. Besides this global version, a local one exist, defined at the level of some node $i$ [7]: $C_i = \frac{\delta_i}{k_i(k_i-1)/2}$, where $k_i$ is the degree of $i$, and $\delta_i$ the number of triangles containing this node. The denominator corresponds to the number of combinations of two neighbors of $i$, in other words: the number of connected triads centered on $i$. The ratio can therefore be interpreted as the probability of finding a direct connection between two neighbors of $i$. The local transitivity can be averaged over the whole network to obtain a global measure. Real-world networks are characterized by a high transitivity, whatever the considered version [2].

Transitivity and community structure are frequently jointly observed in real-world networks. Let us consider for instance the comparative study conducted in [15]. The authors classify networks depending on the systems they model, and analyze their community structures. According to our processing, the transitivity values associated to these community-structured networks are significantly higher than for same-sized random networks, by several orders of magnitude and for all considered classes.

The relationship between transitivity and community structure may seem trivial at first. Intuitively, a high transitivity appears to be the natural consequence of a community structure: links are concentrated in communities and should therefore form many triangles. Reciprocally, it seems a high transitivity indicate the links are form clusters, and therefore communities. However, it is relatively easy to find counter-examples to refute these propositions. First, consider a network whose communities are fully connected multipartite networks: the community structure can be very strong, with dense communities, but the transitivity is nevertheless zero. One could alternatively consider communities taking the form of connected stars, for the same result. Second, consider a fully connected network: the transitivity is maximal, but there is no community structure (just a single community).

To avoid this kind of situation, we based our analysis on randomly generated networks with realistic properties. When possible, we selected generative models able to mimic the topological properties consensually considered to be present in real-world networks.

# 4   Methods

The empirical approach we adopted to study the relationship between community structure and transitivity is two-stepped. First, we generate artificial realistic networks with controllable community structure and analyze how changes in the community structure affect the transitivity. Second, we use two different models able to generate transitive networks, and analyze how changes in the transitivity affect the community structure. The identification of the community structures is performed by applying two different and complementary algorithms. In this section, we describe all three generative models, and summarize the principle of both community detection algorithms.

## *4.1  Community Structure Model*

To generate networks possessing a community structure, we used a modified version of the LFR model [10]. This model applies a three-stepped generative process based on the use of a more basic model, i.e. one not supposed to produce a community structure. First, the basic model is used to generate an initial network. Second, virtual communities are randomly drawn so that their sizes follow a power law distribution. Third, an iterative process takes place to rewire certain links, in order to make the community structure appear while preserving the degree distribution of the initial network.

The strength of the community structure is controlled by a specific parameter called the *mixing coefficient $\mu$*. This parameter allows us to produce networks with various community structure strengths and analyze how this affects the transitivity. The mixing coefficient represents the desired average proportion of links between a node and nodes located outside its community, called inter-community links. Consequently, the proportion of intra-community links is $1-\mu$.

By construction, the LFR model guaranties to obtain power law-distributed community sizes, which is a property present in community-structured real-world networks [10]. Since the degree distribution is preserved during the rewiring step of the generative process, the rest of the topological properties depend mainly on the basic model used at the first step. The original LFR process relies on the Configuration Model (CM) [16], which is able to produce networks with a specified degree distribution. In LFR, it was used to obtain a power law-distributed degree, also a well identified feature of many real-world networks [2]. To detect any potential effect the basic model could have on the transitivity measured in the final networks, we selected two alternatives to the CM, both able to produce scale-free networks too. Barabási–Albert's model (BA) [17] implements a completely different, more realistic, generative process based on preferential attachment. The Evolutionary Preferential Attachment model (EV) [18] is a variant of BA able to produce networks with a higher transitivity.

## *4.2 Transitive Models*

We used two different models to study the effect of the transitivity on the community structure. We first selected a model by Newman [11] (NM), which could be considered as an adaptation of the CM able to produce networks with a controlled transitivity. Instead of specifying the degree $k_i$ of each node $i$ like in the CM, one has to define both the number of single links $s_i$ and the number of distinct triangles $t_i$ attached to the node. In other words, a distinction is made between the links depending on their belonging to a triangle. Both are mutually exclusive, meaning one link is either a single link or appears in only one triangle. In the end, the total degree is $k_i = s_i + 2t_i$. For our study, we wanted to obtain scale-free networks for matters of realism, and we therefore needed to control $k_i$. We consequently introduced in our implementation of NM a parameter called transitivity coefficient $\tau \in [0,1]$, in order to control the proportion of the degree dedicated to triangles (vs. single links). Let [ ] denote the round function, then we have $t_i = [\tau k_i/2]$ and $s_i = [(1-\tau)k_i]$.

The main advantage of NM is it allows artificially changing the transitivity of the generated networks. However, for our study, it also has an important limitation: the obtained transitive structure is not very realistic. Indeed, the created triangles are all distinct, i.e. they cannot share more than one node. Put differently, it is not possible for them to have a common side. This also limits the transitivity (both the global and local versions). The maximal local transitivity some node $i$ can reach is $1/(k_i - 1)$ when $s_i = 0$.

In order to overcome this disadvantage, we developed our *Highly Transitive* model (HT). It is able to randomly generate networks with both a specified degree distribution and a high transitivity. The process starts with a ring network, in order to avoid isolated nodes or components in the final network. Links are then randomly added while respecting the desired degree distribution and favoring the connection of nodes with common neighbors (in order to increase the transitivity).

Our model allows obtaining networks whose transitivity is much higher than in NM networks. However, we are not able to control it with a parameter like we did for NM. Both models are therefore complementary: NM allows us to test for the effect of various level of transitivity, even if the maximal transitivity obtained is not very high (greater than in random networks though, so still realistically high). HT allows us to test for the effect of a very high transitivity on the community structure.

## *4.3 Community Detection*

In the first part of our experiment, the community structure of the generated networks is known, because it is defined by construction. However, this is not the case in the second part, and we therefore need to identify it. For this purpose, we used two recent algorithms: Louvain [19] and Infomap [20].

Louvain (LV) is an optimization algorithm proposed by Blondel et *al*. [19]. It uses a two-stepped hierarchical agglomerative approach. During the first step, the algorithm performs a greedy optimization of the modularity (cf. section 2 ) to identify small communities. During the second step, it builds a new network whose nodes are the communities found during the first step. In this new network, the intra-community links are represented by self-loops, whereas the inter-community links are aggregated and represented as links between the new nodes. The process is repeated on this new network, and stops when only one community remains.

Infomap (INP) is an algorithm developed by Rosvall and Bergstorm [20]. The task of finding the best community structure is expressed as a compression problem. The authors want to minimize the quantity of information needed to represent the path of some random walker traveling through the network. The community structure is represented through a two-part nomenclature based on Huffman coding: the first part is used to distinguish communities in the network and the second to distinguish nodes in a community. With a partition containing few inter-community links, the walker will probably stay longer inside communities, therefore only the second level will be needed to describe its path, leading to a compact representation. The authors optimize their criterion using simulated annealing.

As mentioned in section 2 , many different definitions of the concept of community exist. Louvain optimizes directly the modularity, whereas Infomap relies on a completely different definition of what a community is. The first is from far the most widely spread, and the second proved to be very efficient [21]. From this point of view, these two algorithms are complementary, which is why we selected them. This allows us to detect if the community structure induced by a high transitivity favors one definition or the other.

## 5    Results

### *5.1 Effects of Community Structure on Transitivity*

By applying the LFR rewiring process to the three basic models (CM, BA and EV), we generated three different sets of community structured networks. We selected our parameters values based on previous experiments regarding artificial networks generation [10], and descriptions of real-world networks measurements from the literature [2, 22], so that the produced networks were the most realistic possible. Some parameters are common to all three processes: we fixed the size $n = 5000$ and the power law exponent for the community sizes distribution $\beta = 2$, and made the mixing coefficient $\mu$ range from 0.05 to 0.95 with a 0.05 step. Other parameters are model-dependent. CM allows a precise control of the degree, since it is possible to specify the desired power law exponent $\gamma$ for the degree distribution, and the average $\langle k \rangle$ and maximal degrees $k_{max}$. We used the values $\gamma = 3$, $\langle k \rangle \in \{15, 30\}$ and $k_{max} \in \{45, 90\}$. Both other models do not let as much control, and we had to adjust their parameters so that the resulting networks had approximately the same degree-related properties. Preferential attachment does not give any control on $\gamma$, which tends towards 3 by construction. We produced 25 networks for each combination of parameters, and averaged the transitivity, as shown in Fig. 1 (left). Results were very similar for $\langle k \rangle = 15$ and 30, so we only present the latter here, but comments apply to both.

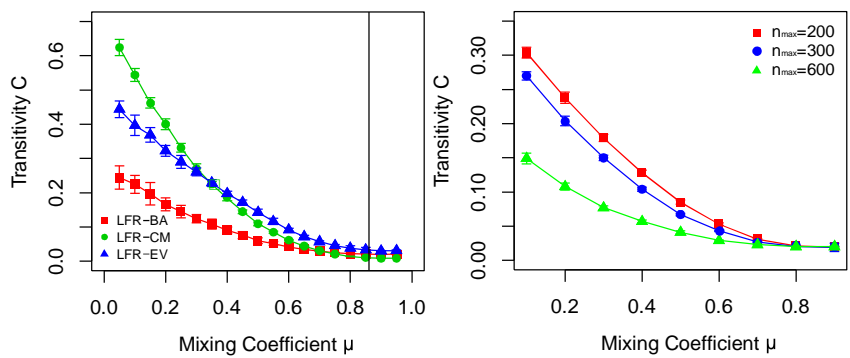

**Fig. 1.** Effect of the mixing coefficient $\mu$ on the transitivity. Each point corresponds to an average over 25 networks generated with $\gamma = 3$ and $\beta = 2$. Left: for each LFR variant, with $n = 5000$, $n_{max} = 700$ and $\langle k \rangle \approx 30$. Right: for several values of $n_{max}$ on LFR-CM, with $n = 1000$ and $\langle k \rangle \approx 15$.

The transitivity of the networks generated by the basic models before rewiring are 0.020, 0.008 and 0.030 for CM, BA and EV, respectively. After rewiring, CM leads to the highest transitivity, with values around 0.6 for $\mu \approx 0$, but it reaches almost zero for $\mu \approx 1$, exhibiting a serious sensitiveness to changes in the community structure strength. Both other models also show a decreasing transitivity when $\mu$ increases, but the range is much smaller, partly because their values for $\mu \approx 0$ are significantly smaller: around 0.25 and 0.45 for BA and EV, respectively. Like CM, their transitivity is close to zero when $\mu \approx 1$. In the literature, real-world networks with a 0.3 transitivity are considered highly transitive [22], so we can state all three models exhibit a realistic transitivity for a small $\mu$ (clearly separated communities). The fact the transitivity decreases when the communities become more and more difficult to discern, for all three models, supports the assumption that a realistic community structure causes a high transitivity.

Besides its strength, a community structure can be characterized by its community size distribution. For realism matters, we chose a power law with fixed exponent $\beta = 2$, but the practical draw of the community sizes requires specifying the size of the largest community $n_{max}$. In order to study the effect of this limit on the transitivity, we generated another batch of networks with $n = 1000$, $\langle k \rangle = 15$ and $n_{max} \in \{200, 300, 600\}$, the other parameters being the same than before. Transitivity values for different largest community sizes are shown on Fig. 1 (right).

When using a smaller $n_{max}$, the size difference between the smallest and the largest communities decreases, making the community size distribution more homogeneous (or rather: less contrasted, since it still follows a power law). It also affects the number of communities, which decreases when $n_{max}$ increases: the numbers of communities are 40, 30 and 15 for $n_{max} = 200, 300$ and $600$, respectively.

It turns out the transitivity measured on the obtained networks decreases when $n_{max}$ increases. In other words, the number of triangles increases when there are less communities, with more similar sizes. This makes sense considering the links constituting triangles have more chance to fall between communities when there are more of them, especially if they are smaller. This is confirmed by the fact the observed effect is stronger for clearly separated communities ($\mu \approx 1$).

## *5.2 Effects of Transitivity on Community Structure*

We specified the parameter values for our HT model so that they were the most similar possible to what was used with LFR. We consequently generated 25 networks with $n = 5000$ and $\langle k \rangle \in \{5, 15, 30\}$, and a power law-distributed degree ($\gamma = 3$). We obtained an average transitivity of 0.5, 0.45 and 0.3 for $\langle k \rangle = 5, 15$ and 30, respectively. This is consistent with the values observed in real-world networks. Both community detection algorithms return modularities close to 0.90, 0.72 and 0.74, respectively, indicating a strong community structure. This observation support the assumption a high transitivity allows obtaining a community structure.

As mentioned before, on the one hand NM does not reach a very high density, but on the other hand it can control it through the transitivity coefficient $\tau$, allowing to analyze how changes in this parameter affects the community structure. It is therefore complementary to our model. Because of the local transitivity limit mentioned in section 4.2, we had to use different parameters (compared to HT) to obtain a relatively high transitivity. We generated 6 networks with $n = 1000$, $\langle k \rangle = 5, 10$, and made $\tau$ range from 0 to 1 with 0.1 steps. Although sparse, the generated networks are connected.

We first focus on the networks obtained for $\langle k \rangle = 5$. Fig. 2 (left) shows how the modularity obtained by both community detection algorithms varies in function of $\tau$. They differ in the amplitude of the measured modularity, which is higher for Louvain than for Infomap. This might be due to the fact Louvain directly optimizes this criterion. However, and more importantly, the trend is the same for both algorithms: the detected community structures get stronger when the transitivity increases. This is particularly true when $\tau > 0.4$. It turns out below this value, the actual transitivity does not change very much ($C < 0.05$), as shown in Fig. 2 (right), certainly due to the rounding performed during the generative process (cf. section 4.2).

The highest modularity is obtained for $\tau = 1$, i.e. when most links are used to create triangles. However, because of the model characteristics, this does not necessarily translates into a very high transitivity value, as shown in Fig. 2 (right).

More surprisingly, even the smallest modularity values (close to 0.5), obtained for $\tau = 0$, are still considered as large in the literature, and reveal a clear community structure. The networks generated for $\tau = 0$ are not supposed to contain many triangles (only those obtained by chance, i.e. a negligible number [11]), as confirmed by the measured transitivity ($C < 0.05$). This indicates a high transitivity is not a prerequisite to the existence of a strong community structure, at least when considering the definition implemented by the modularity measure.

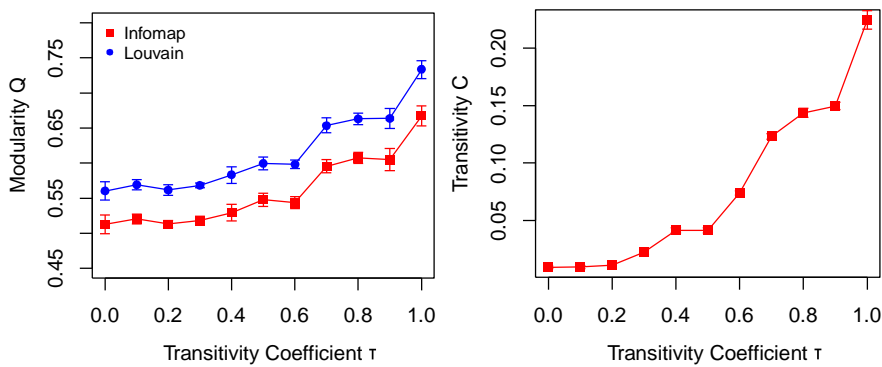

**Fig. 2.** Effect of the transitivity coefficient on the modularity. Each point corresponds to an average over 6 networks generated by NM with $n = 1000$ and $\langle k \rangle = 5$.

For the denser networks ($\langle k \rangle = 10$), the evolution of both the transitivity and modularity are similar to what was observed with $\langle k \rangle = 5$. However, as mentioned before, due to the local transitivity limit present in NM, the transitivity reaches a much lower value of only 0.12: this cannot be considered as high. The modularity is also lower, ranging from 0.33 to 0.41, however these values are still considered as relatively high, even those obtained when $\tau$ is close to zero. This confirms our previous remark regarding the coexistence of both a low transitivity and a significant community structure.

## 6   Conclusion

In this study, we took advantage of several generative models to investigate the relation between the community structure and the transitivity of complex networks. We first applied three variants of the LFR model [10] to generate artificial networks with known community structures. We observed similar results for all three variants: the rewiring process allowing the community structure to appear also causes a large increase in the transitivity. Moreover, the obtained transitivity is directly affected by the strength of the community structure and the distribution of the community sizes. So for this model, transitivity seems to be an offspring of community structure. Secondly, we used two models HT and NM [11] to generate transitive networks. The first, designed by us, produces a very high transitivity, but cannot control it. The transitivity is clearly lower with the second, but a specific parameter allows controlling it. Besides this point, the models are also complementary in the sense they produce networks with very different topologies. We used two state-of-the-art algorithms, Louvain [19] and Infomap [20], to identify the community structures in the generated networks. It turns out the strength of the modularity structure, expressed in terms of modularity, increases with the transitivity, for both generative models and according to both community detection algorithms. This also supports our point concerning the relationship between community structure and transitivity. More surprisingly, according to the obtained modularity, the networks with almost zero transitivity also have a clear (although not as strong) community structure. For NM, it therefore seems the transitivity affects the community structure strength, but is not a prerequisite.

Our main contribution was to study the relationship between community structure and transitivity, which, although intuitively trivial, was not objectively analyzed before. For this purpose, we developed a new random generative model able to produce highly transitive networks with a desired degree distribution. We also modified the other models used in this article, in order to adapt them to our objectives. Our work can be extended in various ways. It would be possible to develop our model, in order to generate more realistic networks, and allow controlling the transitivity. We could also use alternative models, for the production of both community structure and controlled transitivity, in order to ensure our results are not model-dependent. The quality and nature of the community structures could be assessed in a deeper way, through various additional tools like community profile [23] or some alternative to the modularity [12]. There also are generalized versions of the transitivity, dealing with cycles of higher order. But a more important point would be to characterize the *nature* of the relationship between transitivity and community structure. Complementarily to our empirical study, an analytical work would allow identifying the necessary and/or sufficient conditions for the existence of a community structure. This might require to consider other topological properties, especially the network density and degree distribution.

## References

1. Newman, M.E.J., Girvan, M.: Finding and evaluating community structure in networks. Physical Review E 69, - (2004)
2. Newman, M.E.J.: The structure and function of complex networks. SIAM Review 45, 167-256 (2003)
3. Pastor-Satorras, R., Rubi, M., Diaz-Guilera, A., Barabási, A.-L., Ravasz, E., Oltvai, Z.: Hierarchical Organization of Modularity in Complex Networks. Statistical Mechanics of Complex Networks, Vol. 625, 46-65. Springer Berlin / Heidelberg (2003)
4. Clauset, A., Moore, C., Newman, M.E.J.: Hierarchical structure and the prediction of missing links in networks. Nature 453, 98-101 (2008)
5. Liu, Z.H., Bambi: Epidemic spreading in community networks. Europhysics Letters 72, 315-321 (2005)

6. Wang, G.-X., Qin, T.-G.: Impact of Community Structure on Network Efficiency and Communicability. Intelligent Computation Technology and Automation (ICICTA), 2010 International Conference on, Vol. 2 (2010) 485-488
7. Watts, D., Strogatz, S.H.: Collective dynamics of 'small-world' networks. Nature 393, 409-410 (1998)
8. Jin, E.M., Girvan, M., Newman, M.E.J.: Structure of growing social networks. Phys. Rev. E 64, 046132 (2001)
9. Boguña, M., Pastor-Satorras, R., Diaz-Guilera, A., Arenas, A.: Models of social networks based on social distance attachment. Phys. Rev. E 70, 056122 (2004)
10. Lancichinetti, A., Fortunato, S., Radicchi, F.: Benchmark graphs for testing community detection algorithms. Phys Rev E 78, 046110 (2008)
11. Newman, M.E.J.: Random Graphs with Clustering. Phys Rev Lett 103, - (2009)
12. Fortunato, S.: Community detection in graphs. Physics Reports 486, 75-174 (2010)
13. Fortunato, S., Barthelemy, M.: Resolution limit in community detection. PNAS USA 104, 36-41 (2007)
14. Luce, R.D., Perry, A.D.: A method of matrix analysis of group structure. Psychometrika 14, 95–116 (1949)
15. Lancichinetti, A., Kivelä, M., Saramäki, J., Fortunato, S.: Characterizing the Community Structure of Complex Networks. PLoS ONE 5, e11976 (2010)
16. Molloy, M., Reed, B.: A critical point for random graphs with a given degree sequence. Random Structures and Algorithms 6, 161-179 (1995)
17. Barabasi, A., Albert, R.: Emergence of scaling in random networks. Science 286, 509 (1999)
18. Poncela, J., Gomez-Gardeñes, J., Florıa, L.M., Sanchez, A., Moreno, Y.: Complex Cooperative Networks from Evolutionary Preferential Attachment. PLoS ONE 3, e2449 (2008)
19. Blondel, V.D., Guillaume, J.-L., Lambiotte, R., Lefebvre, E.: Fast unfolding of communities in large networks. J Stat Mech P10008 (2008)
20. Rosvall, M., Bergstrom, C.T.: Maps of random walks on complex networks reveal community structure. PNAS 105, 1118 (2008)
21. Orman, G.K., Labatut, V., Cherifi, H.: Qualitative Comparison of Community Detection Algorithms. Communications in Computer and Information Science 167, 265-279 (2011)
22. da Fontoura Costa, L., Oliveira Jr., O.N., Travieso, G., Rodrigues, r.A., Villas Boas, P.R., Antiqueira, L., Viana, M.P., da Rocha, L.E.C.: Analyzing and Modeling Real-World Phenomena with Complex Networks: A Survey of Applications. arXiv 0711.3199, (2008)
23. Leskovec, J., Lang, K.J., Dasgupta, A., Mahoney, M.W.: Statistical Properties of Community Structure in Large Social and Information Networks. In: WWW, ACM, Beijing, CN (2008)